\documentstyle[PASJadd,epsf]{PASJ95}
\draft

\begin{document}
\setcounter{page}{1}

\title{Gravitational-Wave Radiation from Magnetized Accretion Disks}

\author{Shin {\sc Mineshige}, Takashi {\sc Hosokawa} \\
{\it Yukawa Institute for Theoretical Physics, Kyoto
University, Sakyo-ku, Kyoto 606-8502}
\\
{\it  minesige@yukawa.kyoto-u.ac.jp}
\\
and 
\\
Mami {\sc Machida} and Ryoji {\sc Matsumoto}
\\
{\it Department of Physics, Chiba University, 1-33 Yayoi-Cho, 
  Inage-ku, Chiba 263-8522} 
}

\abst{The detectability of gravitational wave (GW) radiation from
accretion disks is discussed based on various astrophysical contexts.
In order to emit GW radiation, the disk shape should lose
axial symmetry.  We point out that a significant deformation is plausible
in non-radiative hot accretion disks because of enhanced magnetic activity,
whereas it is unlikely for standard-type cool disks.
We have analyzed the 3D magnetohydrodynamical (MHD) simulation data
of magnetized accretion flow,
finding non-axisymmetric density patterns.  
The corresponding ellipticity is $\epsilon \sim 0.01$.
The expected time variations of GW radiation are overall chaotic,
but there is a hint of quasi-periodicity.
GW radiation has no interesting consequence, however,
in the case of close binaries, because of very tiny disk masses.
GW radiation is not significant, either, for AGN
because of very slow rotation velocities.
The most promising case can be found in gamma-ray bursts or supernovae,
in which a massive torus (or disk) with a solar mass or so may be
formed around a stellar-mass compact object
as the result of a merger of compact objects, or by the fallback 
of exploded material towards the center in a supernova.
Although much more intense GW radiation is expected 
before the formation of the torus,
the detection of GW radiation in the subsequent accretion phase
is of great importance, since it will provide a good probe
 to investigating their central engines.
}

\kword{accretion, accretion disks --- black holes --- galaxies: Seyfert
--- gamma-ray bursts --- gravitational wave}

\maketitle

\section{Introduction}

Gravitational wave (GW) radiation is expected to be detected
within 10 years or so, thereby opening a new window
to probe extremely high-density objects and the early universe 
(see, e.g. Thorne 1987, 1997; Blair 1991; Tsubono et al. 1997).
When detected, the impact will be enormous and the understanding of
astrophysical objects should inevitably undergo revolutionary development.
It is thus of great importance to discuss at this moment what kinds of
objects can emit GW radiation.  In the present study,
we consider if it is feasible to detect GW radiation from 
accretion disks in general contexts.

At first glance, it is unreasonable to expect GW radiation from
accretion disks, since,
while the generation of GW requires the presence of quadruple moment,
the usual disk models are constructed under the assumption of axisymmetry.
Actually, the
axisymmetric assumption is thought to be quite generally satisfied
(see, e.g., Kato et al. 1998 for a review of various disk models).
Then, how and under what circumstances can non-axisymmetry arise?

A non-axisymmetric disk structure is often discussed in the
context of accretion disks in close binary systems. 
Since the shape of the Roche lobe
is not totally spherically symmetric, 
we expect a disk there to lose axial symmetry,
especially when the disk size is large, close to the size of the Roche lobe
(Paczy\'nski 1977).  Another good indication of non-axisymmetric structure
is frequently discussed based on hydrodynamical simulations; e.g.
Sawada et al. (1986).  
Note that the presence of spiral patterns on a disk was discovered 
through the observing technique of Doppler tomography
(Steeghs et al. 1997).
As we will see later, however, because disks in close binary systems
have tiny mass, the production of strong GW radiation is unlikely.

Alternatively, strong magnetic fields, if they ever exist,
could greatly modify the disk structure.  This
is the subject of the present study.  
As was first discussed by Shakura and Sunyaev (1973)
and later demonstrated by many authors through MHD simulations
(Matsumoto 1999; Stone et al. 1999; Machida et al. 2000; Hawley et al. 2001),
the magnetic energy can be amplified by the
number of MHD instabilities together with differential rotation
up to the value of $p_{\rm mag}/p_{\rm gas} \simeq$ 0.01--1
(e.g. Machida et al. 2000).  
The corresponding viscosity parameter is $\alpha \simeq 0.01 - 0.1$.
This is just in the range that
we require to account for the observations of
dwarf-nova outbursts (Cannizzo 1993) and
X-ray nova eruptions (Mineshige, Wheeler 1989).

In standard-type disks,
however, it is hard to believe a large influence of magnetic fields
on the shape of the disk, since the magnetic energy is, at most,
comparable to the internal energy of the gas, which is
much less than the kinetic (rotational) energy of the gas
as a consequence of efficient radiative cooling of the disk material.
In other words, the magnetic pressure cannot overcome the gravitational force
in such radiation-dominated accretion flow.
In advection-dominated regimes (see Kato et al. 1998 for a review), 
in contrast, the internal energy of gas can be comparable to its kinetic
energy and potential energy because of a low radiation efficiency.
We can then expect a large influence of magnetic fields on the
dynamics of accretion disks.

Another argument to support this idea comes from the observed
complex variability commonly observed in many black-hole candidates
during their hard state.  In that state, the spectra are hard,
and a hot accretion flow model (ADAF model) can fit the observations
(Ichimaru 1977; Narayan et al. 1996).  
Although its origin is not yet established, many authors suggest that
the variability could be caused by a sporadic release of magnetic energy 
triggered by magnetic reconnection and flares (Takahara 1979; 
Galeev et al. 1979).
Without large magnetic field energy, it is difficult to account for
substantial variations.
This idea is consistent with the absence of large fluctuations
during the soft state, when a standard-type disk seems to be present.
It then follows that a
non-axisymmetric disk structure may be generated by
magnetic fields in cooling-inefficient regimes (i.e.
adiabatic regimes) of disk accretion.
In fact,  the global MHD simulation of a disk exhibits a
spatial inhomogeneous structure (Kawaguchi et al. 2000).
Finally, note that large fluctuations are also observed in
the high-luminosity state.  Magnetic activities also seem to be
enhanced in such a state (e.g. Mineshige et al. 2000).

We, here, consider GW radiation from
accretion disks, the shapes of which are possibly influenced 
by large magnetic fields or other effects.
In section 2 we calculate the moments of inertia
by using the MHD simulations data and see to what extent a
deviation from an axisymmetric disk can be expected.
We then discuss the detectability of GW radiation in 
various astrophysical contexts.
The final section is devoted to discussion.

\section{Inhomogeneous Density Structure Created by Magnetic Fields}
Here, we analyze 3D MHD simulation data newly calculated by
Machida and Matsumoto (2002).  
They started the simulation with a torus (Okada et al. 1989) 
threaded by weak toroidal fields (see Machida et al. 2000)
in a pseudo-Newtonian potential (see similar calculations 
by Hawley, Krolik 2001).
The size of the calculation box was taken to be 100 $r_{\rm g}$
(with $r_{\rm g}$ being the Schwarzschild radius)
and the mass within the last stable circular orbit at $3~r_{\rm g}$
was removed as an inner boundary condition.  
We have confirmed that
magnetic fields are amplified up to the value of 
$p_{\rm mag}/p_{\rm gas} \simeq 0.1$ within $\sim 50$ rotation periods
at the reference radius of $r_0 = 50~r_{\rm g}$
(see Machida, Matsumoto 2002 for more details).

\begin{figure}[t]
\epsfxsize\columnwidth \epsfbox{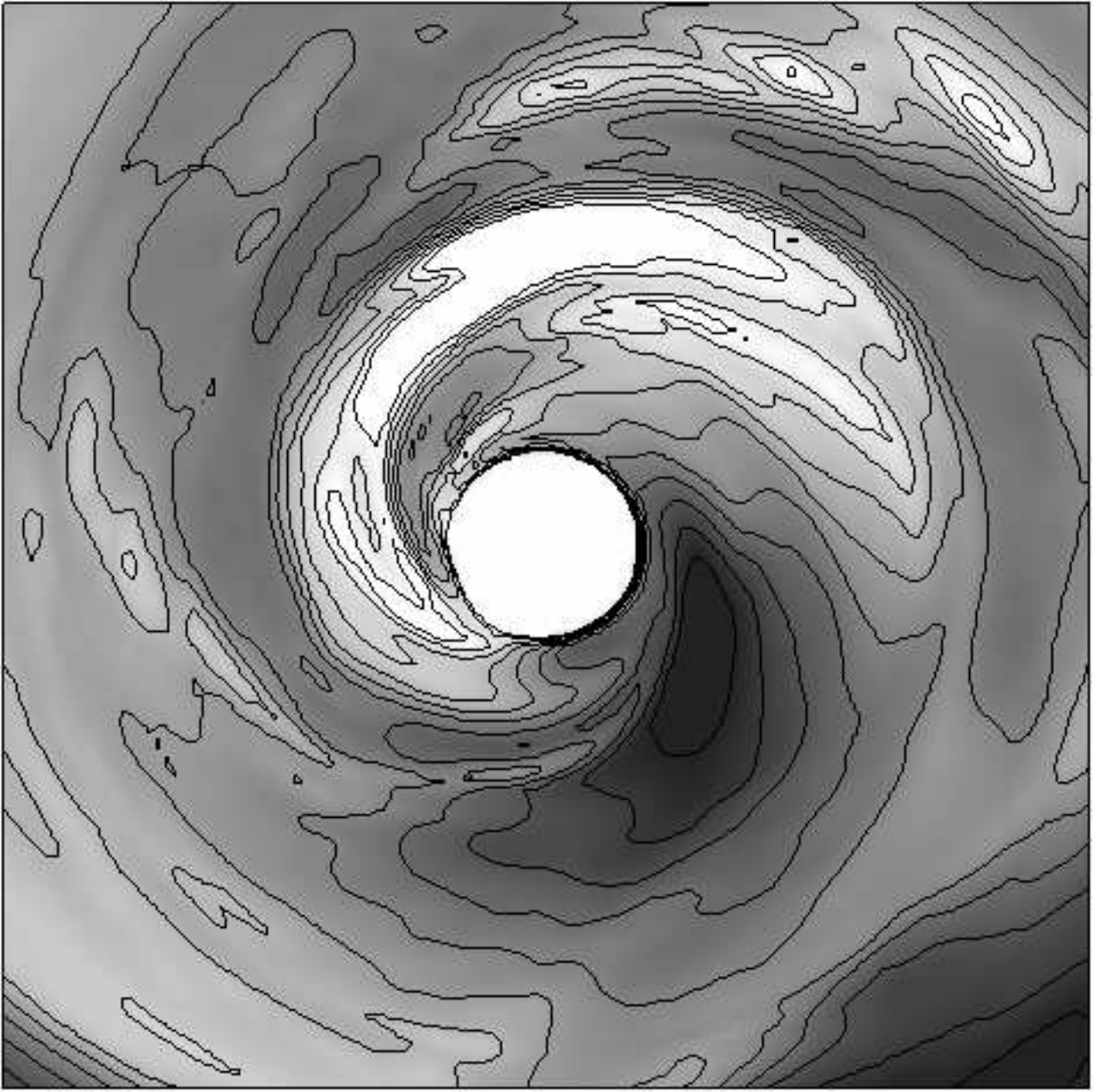}
\caption{
Snapshot of the density contours of a magnetized disk
on the equatorial plane [at a calculation time of
$t=32070 (r_{\rm g}/c)$, see Machida, Matsumoto 2002].
This figure covers the region of 20 $r_{\rm g} \times 20~r_{\rm g}$.
The spacing between each contour is $\Delta\log\rho = 0.05$.
}
\label{frac}
\end{figure}

Figure 1 represents a snapshot of the density contours on the
disk equatorial plane. A sort of spiral pattern (with the azimuthal
number of the spiral being $m = 1$) is clear in this plot.
As time goes on, this pattern shows a rotation around the center,
thus producing a non-zero ellipticity.  However,
the pattern is not the same in the rotational frame due to
the presence of differential rotation and magnetic energy dissipation
by magnetic reconnection.  

To evaluate the strength of GW radiation, we next
calculate the moments of inertia using the MHD simulation data.
Here, we use cylindrical coordinates, $(r,\varphi,z)$, 
with the $z$-axis being taken as the rotational axis. Then,
the moments of inertia are written as
\begin{equation}
   I_{xx} \equiv \int \rho(r,\varphi,z) x^2 r dr d\varphi dz
\end{equation}
and
\begin{equation}
   I_{yy} \equiv \int \rho(r,\varphi,z) y^2 r dr d\varphi dz
\end{equation}
with $x \equiv r\cos\varphi$ and $y \equiv r\sin\varphi$.
Since the spiral density pattern rotates with time,
such a magnetized disk can certainly produce GW radiation,
which is rather chaotic, as is shown in figure 2.
At the same time, we also notice a hint of quasi-periodic variations
on timescales of $\sim 200~r_{\rm g}/c$ 
in the variations of the ellipticity ($\epsilon$) 
and moments of inertia ($I_{xx}$ and $I_{yy}$).

The different behavior of $I_{xx}$ and $I_{yy}$ yields an ellipticity, 
which is calculated by
\begin{equation}
    \epsilon \equiv \frac{I_{xx}-I_{yy}}{I_{xx}+I_{yy}}.
\end{equation}
As is well known, non-axisymmetric disks (with $\epsilon\neq 0$)
can emit GW radiation. In the next section
we estimate how much energy can be extracted by the GW radiation
and how large an amplitude of GW radiation is expected for a simple case of
a non-axisymmetric disk with an ellipticity of $\epsilon \sim 0.01$.
It might be noted here that, precisely, 
although we need to sum up all of the contributions from each mass particle 
moving around the center, for the purpose of 
order-of-magnitude discussion presented in this paper,
it will suffice to use the formula for an elliptical mass distribution.

\begin{figure}[t]
 \epsfxsize\columnwidth \epsfbox{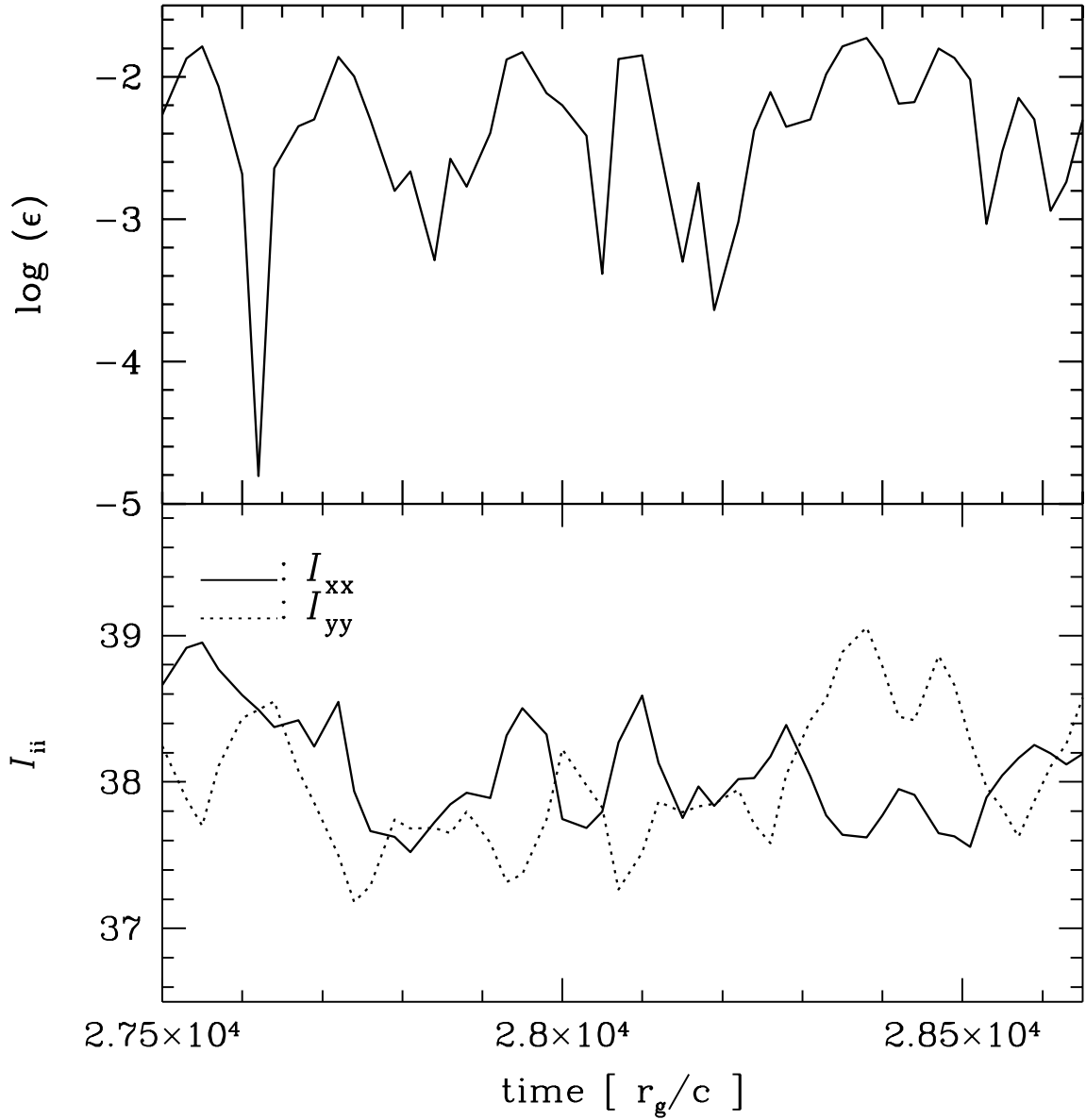}
\caption{
Time variations of the ellipticity (upper panel),
and those of moments of inertia, $I_{xx}$ (by the solid lines)
and $I_{yy}$ (by the dashed lines), respectively (lower panel).
The units of the time and moment of inertia are 
$r_{\rm g}/c$ ($\approx 3\times 10^{-5}$ s for the central mass of 
$M_* = 3~M_\odot$) and $(M_{\rm disk}/M_\odot)(r_{\rm disk}/10^7{\rm cm})^2$, 
respectively.  We find $\epsilon \sim 0.01$.
}
\label{frac}
\end{figure}

\section{Possible GW Radiation from Accretion Disks}
\subsection{Basic Formulae}
In the previous section, we have seen how a non-axisymmetric disk
structure is created by global magnetic fields.  We expect
about a few percent of inhomogeneity in the magnetized disk.
We then proceed the discussion and consider
what the most ideal targets would be.

The energy loss rate due to the GW radiation is expressed by
\begin{equation}
   {\dot E}_{\rm gw} = -\frac{32}{5}\frac{G}{c^5}I^2\epsilon^2\Omega^6,
\end{equation}
whereas the expected amplitudes are on the order of
\begin{equation}
   h_+ = -\langle h \rangle (\cos^2\theta+1)\cos(2\Omega t-2\varphi_0)
\end{equation}
and
\begin{equation}
 h_\times = -2\langle h\rangle \cos\theta \sin(2\Omega t-2\varphi_0)
\end{equation}
with a coefficient of
\begin{equation}
  \langle h\rangle \equiv (2\epsilon GI\Omega^2)/(c^4d),
\end{equation}
and $\varphi_0$ being a constant.  Note that the effective amplitude
is $h_{\rm eff} = \sqrt{N}\langle h\rangle$ with $N$
being the number of the cycles with regard to the quasi-periodicity
during the observing time.

\subsection{Case of Binary Accretion Disks}
For disks around neutron stars or black holes in close binary systems, 
typical parameters are a disk mass of $M_{\rm disk}\sim 10^{-8} M_\odot$,
a disk size of $r_{\rm disk} \sim 10^{10}$cm, and an orbital period
at the disk rim of 
\begin{equation}
   P = 2\pi\sqrt{\frac{r_{\rm disk}^3}{GM_*}} 
        \sim 300~
        \left(\frac{m_*}{3}\right)^{-1/2} r_{10}^{3/2} [{\rm s}],
\end{equation}
where $m_* \equiv M_*/M_\odot$
with $M_*$ being the mass of the central object,
and $r_{10}\equiv r_{\rm disk}/10^{10}$cm.
We then find
\begin{equation}
   {\dot E}_{\rm gw} = -4.5 \times 10^{17}
             \left(\frac{m_{\rm disk}}{10^{-8}}\right)^{2} 
                    \frac{\epsilon_{-2}^2}{r_{10}^{5}} 
                \left(\frac{m_*}{3}\right)^{3} [{\rm erg~s}^{-1}],
\end{equation}
where $m_{\rm disk}\equiv M_{\rm disk}/M_\odot$ and
$\epsilon_{-2} \equiv \epsilon/10^{-2}$.
The timescale of the energy loss from the disk would be
\begin{eqnarray}
\label{t_binary}
 t_{\rm gw} \!\!&\equiv&\!\! \frac{E_{\rm rot}}{|{\dot E}_{\rm gw}|} \cr
            \!\!&\simeq&\!\! 2.8  \times 10^{16}
             \left(\frac{m_{\rm disk}}{10^{-8}}\right)^{-1} 
                \frac{r_{10}^4}{\epsilon_{-2}^2}
                \left(\frac{m_*}{3}\right)^{-2}  [{\rm yr}].
\end{eqnarray}
Since the characteristic timescale greatly exceeds the age of the universe, 
GW radiation is totally unimportant for the evolution of binary disks.
The expected GW amplitude is
\begin{equation}
\label{amp_binary}
  h_{\rm eff} \simeq  1.3 \times 10^{-29}
             \left(\frac{m_{\rm disk}}{10^{-8}}\right)
                    \frac{\epsilon_{-2}}{r_{10}} 
              \left(\frac{m_*}{3}\right)
               \left(\frac{d}{1~{\rm kpc}}\right)^{-1}
\end{equation}
for $N=1000$.
This is far below the current detection limit.  There is no hope to
detect GW radiation from the binary accretion disks.

\subsection{Case of Massive Disks in GRBs or HNe}
Next, we consider a massive disk around a stellar-mass compact object.  
It has been proposed that a massive torus or disk 
(with solar mass or so) is possibly formed around a stellar-mass compact 
object after the merger of compact objects,
a promising possibility to produce gamma-ray bursts
(e.g., Narayan et al. 1992; Kohri, Mineshige 2002).
Likewise, fallback of exploded material towards the center 
may also form a massive torus in a supernova (SN) or hypernovae (HN)
(Woosley 1993).  For recent reviews of GRBs, see
Piran (2000); M\'esz\'aros (2001), and references therein.
We thus expect much more intense GW radiation in these cases
than in the case of close binaries.

The typical parameters are 
$M_{\rm disk}\sim M_\odot$ and $r_{\rm disk} \sim 10^{7}$cm;
and the orbital period at the disk rim is
\begin{equation}
   P \sim 0.01 ~
              \left(\frac{m_*}{3}\right)^{-1/2} r_7^{3/2} [{\rm s}],
\end{equation}
with $r_7 \equiv r_{\rm disk}/10^7$ cm.
That is, the modulation frequency is $f \equiv 2/P \sim 200$ Hz or so.
From the formula we find the energy loss rate to be
\begin{equation}
   {\dot E}_{\rm gw} = -4.5 \times 10^{48}
                \frac{m_{\rm disk}^2\epsilon_{-2}^2}{r_{7}^{5}} 
                \left(\frac{m_*}{3}\right)^{3} [{\rm erg~s}^{-1}].
\end{equation}
The timescale of the energy loss from the disk is estimated to
\begin{equation}
\label{t_GRB}
   t_{\rm gw}  \simeq 10^4 
                 \frac{r_{7}^4}{m_{\rm disk}\epsilon_{-2}^2}
                 \left(\frac{m_*}{3}\right)^{-2} [{\rm s}].
\end{equation}
Since the lifetime of a disk in one GRB is several tens to thousands of seconds,
at longest, GW radiation is not very important.  Note, however,
that more deformation of the disk shape is plausible if the
massive torus is unstable against self-gravity.
Then, GW radiation may affect the secular evolution of the disk.
The effective amplitude would be
\begin{equation}
\label{amp_GRB}
   h_{\rm eff} \simeq 1.3 \times 10^{-22} \sqrt{N_3}
                  \frac{m_{\rm disk}\epsilon_{-2}}{r_{7}} 
                 \left(\frac{m_*}{3}\right)
                 \left(\frac{d}{10~{\rm Mpc}}\right)^{-1},
\end{equation}
where $N_3 \equiv N/1000$ and we took
$N \sim$ 10 s / 0.01 s = $10^3$.
This is marginally detectable by the LIGO, as long as a GRB
occurs within a distance of $\sim$ 10 Mpc.  Hence, 
there is a possibility of detecting GW radiation from hypernovae or
GRBs, although much more enhanced GW radiation is expected 
in a violent phase (i.e. SN explosion or tidal destruction of a star by
a black hole before a merging) preceding the formation of a massive disk
(e.g. Yamada, Sato 1995; Shibata, Uryu 2000).
Nevertheless, the detection of GW radiation from the disk accretion phase
is crucial, since it would continue for the burst durations,
thus providing an essential clue to understanding
what is actually going on at the center of SNe and GRBs.

\subsection{Case of AGN Accretion Disks}
Finally, let us consider accretion disks in AGN.
There is a convenient scaling law describing the mass distribution
and time scales of AGN accretion disks.
Since the surface density does not strongly depend on the black-hole mass,
as long as the radial distance to the center is expressed in terms of
the Schwarzschild radius ($r_{\rm g} \propto M_{\rm BH}$)
and the mass-accretion rate is scaled by the Eddington rate
($L_{\rm Edd}/c^2$ with $L_{\rm Edd}$ being the Eddington luminosity),
we have the following approximate scaling laws:
\begin{equation}
    M_{\rm disk} \sim \pi r_{\rm disk}^2 \Sigma \propto M_{\rm BH}^2
\end{equation}
and
\begin{equation}
   I \sim M_{\rm disk}r_{\rm disk}^2 \propto M_{\rm BH}^4.
\end{equation}
These scalings exactly hold, if the disk is radiation pressure-dominated
and if the disk size is proportional to the black-hole mass.
Numerically, we thus estimate 
\begin{equation}
  M_{\rm disk} \sim 10^{-8} \left(\frac{m_*}{3}\right)^2M_\odot
                \approx 10^{7}~ \left(\frac{m_*}{10^8}\right)^2 M_\odot
\end{equation}
and
\begin{equation}
   r_{\rm disk} \sim 10^{10}\left(\frac{m_*}{3}\right)~[{\rm cm}]
                \approx 0.1~ \left(\frac{m_*}{10^8}\right) [{\rm pc}].
\end{equation}
Likewise, the orbital periods and angular frequencies can be scaled as 
\begin{equation}
  P \sim 10^{10}\left(\frac{m_{\rm BH}}{3}\right)~[{\rm s}] \propto M_{\rm BH}
\end{equation}
(with $m_{\rm BH} \equiv M_{\rm BH}/M_\odot$)
and, hence, $\Omega \propto M_{\rm BH}^{-1}$.
Therefore, we find
\begin{equation}
\label{t_AGN}
   t_{\rm gw} \propto I\Omega^4 \propto M_{\rm BH}^0,
\end{equation}
that is, the characteristic timescale is mass-independent and
thus again exceeds the age of the universe [see equation (\ref{t_binary})].
(The timescale may depend on the black-hole mass,
 if the disk size is not proportional to the black-hole mass;
but, even for such a case it is highly unlikely that GW radiation 
significantly affects the time evolution of AGN disks.)

How about the effective amplitudes?
By using equation (\ref{amp_binary}), we easily obtain
\begin{equation}
   h_+ \propto I\Omega^2/d \propto M_{\rm BH}^2,
\end{equation}
leading to the estimation
\begin{equation}
\label{amp_AGN}
  h_{\rm eff} 
       \simeq  1.3 \times 10^{-18} 
                \sqrt{N_3}{\epsilon_{-2}}
               \left(\frac{m_{\rm BH}}{10^8}\right)^2
               \left(\frac{d}{10~{\rm Mpc}}\right)^{-1}.
\end{equation}
Thus, the amplitude is not small, as expected, because of a massive disk.
However, the frequency is only
$f = 2/P \sim 2\times 10^{-10}$ Hz; that is, there are no
GW detectors planned which are capable of detecting such extremely
low GW radiation.  

Because the disk is rotating more rapidly in the inner portions, 
there may be a chance of detecting GW radiation from the innermost 
part of the AGN disks.
Let us consider the innermost part with a radius of $r_0$.  Then,
the mass and
the moment of inertial are reduced by factors of $(r_{\rm out}/r_0)^2$
and $(r_{\rm disk}/r_0)^4$, respectively, while the angular frequency is
increased by a factor of $(r_{\rm disk}/r_0)^{3/2}$,
as long as the Kepler rotation law is adopted.
By setting $r_0 \sim 10~r_{\rm g} \sim 10^{-3} r_{\rm disk}$, we obtain
\begin{equation}
\label{amp_AGN2}
   h_{\rm eff} \simeq  1.3 \times 10^{-21}
                \sqrt{N_3}{\epsilon_{-2}}
               \left(\frac{m_{\rm BH}}{10^8}\right)^2
                \left(\frac{d}{10~{\rm Mpc}}\right)^{-1},
\end{equation}
and the frequency is 
\begin{equation}
\label{f_AGN2}
   f \sim 6\times 10^{-6} 
            \left(\frac{m_{\rm BH}}{10^8}\right)^{-1} ~[{\rm Hz}].
\end{equation}
Still, it is difficult to detect GW radiation for 
AGN located at $d \sim 10$ Mpc by LISA.

\section{Discussion and Conclusion}

We have discussed the detectability of GW radiation from accretion disks
based on the assumption that disks are magnetically dominated, and
thus their structure is inhomogeneous, leading to the emergence of
non-axial symmetry.
In the case of close binaries, however, GW radiation is totally
negligible because of their very tiny disk masses.
In the case of AGN, GW radiation is not significant, either,
because of slow rotation velocities.
Instead, we may concentrate on the innermost part of the AGN disks,
which has much faster rotation velocities.
In fact, the inner hot part of AGN disks seems to be
in a non-radiative state, thus producing hard X-ray emission.
Yet, the detection of GW radiation is not very feasible
with the currently planned GW detectors, unless additional mechanisms
work to enhance the disk deformation.

The most probable case may be found in gamma-ray bursts
and supernovae (or hypernovae), in which
 a massive disk (with solar mass or so) is suggested to be formed 
around a stellar-mass compact component.
Note, however, that
the time variations of GW radiation are not predictable;
they are highly chaotic, although there is a hint of quasi-periodicities.
Also note that much more intense GW radiation is expected before
the formation of a massive torus.  Nevertheless, the detection
of GW radiation from such a massive disk is of great importance to
probe the central engines of the GRBs and SNe (HNe).

The accretion disks can emit electro-magnetic (EM) wave radiation as well as
GW radiation.  The former might be even more important 
than the latter, when considering the long-term evolution of the disk,
since whereas the emissivity of GW radiation is proportional to $\Omega^4$, 
that of the EM wave radiation is proportional to $\Omega^2$.
One might think that 
the disk rotation may then be efficiently decelerated by
the EM wave radiation so that significant GW radiation cannot be expected.
However, this does never occur, since the
removal of angular momentum of a disk gas results in faster rotation,
rather than slower rotation, unlike the case of rotating stars.
This is because when the
disk material loses angular momentum, $\ell$ ($ \propto \sqrt{r}$),
the disk gas moves inward because of the reduced centrifugal force 
(which is $\propto \ell^2$),
leading to more rapid rotation with higher angular frequency
(note that $\Omega \propto r^{-3/2}$).
This makes a good contrast with the case of rotating neutron stars, 
in which the removal of angular momentum results in a spin-down of the star.

Further, we find it unlikely that the evolution of the disk is
totally modified by the EM wave radiation, since the magnetic energy,
which produces EM wave radiation, 
is only $\sim$ 1\% of the gravitational energy
(which is equal to the rotational energy by the Virial theorem).
To be more precise, 
 the Alfv\'en wave is more important than EM wave radiation
 in terms of the energy loss from the disk.
The Poynting flux due to Alfv\'en waves propagating
along the poloidal magnetic field is
\begin{equation}
  {\dot E}_{\rm Alf} \sim B^2 v r^2,
\end{equation}
where $v$ is on the order of the rotation velocity
of the footpoint of the poloidal field anchored by the disk; that is
$v = r\Omega$.  On the other hand,
the energy loss due to the EM wave radiation is at most
\begin{equation}
  {\dot E}_{\rm EM} \sim (\Omega^2 B r^3)^2/c^3,
\end{equation}
since the magnetic moment is of the order of $Br^3$ at maximum.
Thus, we have
\begin{equation}
  {\dot E}_{\rm EM} \sim {\dot E}_{\rm Alf} (r\Omega/c)^3 < {\dot E}_{\rm Alf}.
\end{equation}
The characteristic time of the disk evolution due to the
emission of Alfv\'en wave is then
\begin{equation}
   t_{\rm Alf} \sim (GM_*\rho/r) B^{-2} \Omega^{-1} \gg \Omega^{-1},
\end{equation}
[recall $B^2/(GM_* \rho/r) \sim 0.01$].
Note that this seems to be highly underestimated, since the magnetic fields
have bisymmetric spiral patterns (Machida, Matsumoto 2002)
so that the magnetic moment should be much less than what we assumed above.
We can thus safely conclude that the EM effects are not very significant,
and, thus, our argument concerning the GW radiation from the disk 
is not basically altered.

Since different parts of the disk material rotate with different
angular frequencies ($\Omega$),  the actual
 GW radiation could be the superpositions of many modes 
which vary on a variety of periods.  This leads to a reduction in
the effective GW amplitudes estimated in the previous section.
Whether this is the case or not strongly depends on the
magnetic-field distribution in the disk.  
(Such a problem does not occur in the GRB case, since
 only a compact disk has been considered there.)
If magnetic fields are dominant in the innermost region 
(within, say, several hundreds of Schwarzschild radii),
clear spiral patterns are only appreciable in a relatively narrow region.
Unfortunately, however, 
it is still difficult to calculate the long-term evolution of the
magnetized disk with sufficiently large spatial dimension.  Further,
it is currently impossible to simulate
two-phased disk corona structure with proper treatments of
magnetic fields and radiation transfer.
We need to await until it becomes possible to calculate
the coupling between the radiation, magnetic fields, and matter
in accretion flow.

Strong GW radiation from the gamma-ray burst
was discussed by van Putten (2001) in a slightly different context.
He conjectured that magnetic fields with poloidal topology
equivalent to pulsar magnetospheres provide tight coupling between
a Kerr hole and suspended accretion torus, thereby radiating
GW by extracting the spin-energy of the black hole 
in the presence of non-axisymmetry.
In the present study, in contrast, we do not assume the presence of
rapidly spinning black holes, nor tight coupling between the
black hole and the torus, which makes a distinction between his model
and ours.  Our model should work in any disks with hot gas, in general, and
does not require any special condition.

We wish to finally note that
for a massive disk whose mass is comparable to, or exceeds,
that of the central object,
the torus itself becomes gravitationally unstable, thus giving rise
to non-axisymmetry, even without magnetic fields
(e.g. Woodward et al. 1994).  Then, the corresponding $\epsilon$
value will be much increased: say, $\epsilon \sim 1$ in the extreme case of
fragmentation of the disk matter into several pieces.
This may be relevant in the case of a compact massive disk as in GRBs and
HNe and strengthens our conclusion.

\bigskip
\bigskip
We greatly appreciate a number of valuable comments made by an anonymous referee.
We also acknowledge the Yukawa Institute for Theoretical Physics
at Kyoto University, where this work was initiated
during the YITP workshop YITP-W-01-16 
on {\lq\lq}gravitational waves.{\rq\rq}
This work was partially supported by
Japan Science and Technology Corporation (ACT-JST) and
by Grants-in Aid of the Ministry of Education, Culture, Sports, Science 
and Technology of Japan (13640238, SM).
Numerical computations were carried out by using
Fujitsu VPP300/16R at National Astronomical Observatory, Japan
and Yukawa Institute Computer Facility.

\section*{References}
\small
\re Blair, D. G. 1991, The Detection of Gravitational Waves
     (Cambridge: Cambridge University Press)
\re Cannizzo, J. K. 1993, in Accretion Disks in Compact Stellar Systems, 
    ed. J. C. Wheeler (Singapore: World Scientific Publishing), 6
\re Galeev, R., Rosner, R., \& Vaiana, G. S. 1979, ApJ, 229, 318
\re Hawley, J. F., Balbus, S. A., \& Stone, J. M. 2001, ApJ, 554, L49
\re Hawley, J. F., \& Krolik, J. H. 2001, ApJ, 548, 348
\re Ichimaru, S. 1977, ApJ, 214, 840
\re Kato, S., Fukue, J., \& Mineshige, S. 1998, Black-Hole Accretion Disks
    (Kyoto: Kyoto University Press)
\re Kawaguchi, T., Mineshige, S., Machida, M., Matsumoto, R. \& Shibata, K. 2000,
    PASJ, 52, L1
\re Kohri, K., \& Mineshige, S. 2002, ApJ, in press
\re Machida, M., Hayashi, M. R., \& Matsumoto, R. 2000, ApJ, 532, L67
\re Machida, M., \& Matsumoto, R. 2002, ApJL, submitted
\re Matsumoto, R. 1999, Numerical Astrophysics, ed. S. M. Miyama,
    K. Tomisaka, \& T. Hanawa (Dordrecht: Kluwer Academic Publishers), 195
\re M\'esz\'aros, P. 2001, Science, 291, 79
\re Mineshige, S., Kawaguchi, T., Takeuchi, M. \& Hayashida, K. 2000,
    PASJ, 52, 499
\re Mineshige, S., \& Wheeler, J. C. 1989, ApJ, 343, 241
\re Narayan, R., McClintock, J. E., \& Yi, I. 1996, ApJ, 457, 821
\re Narayan, R., Paczy\'nski, B., \& Piran, T. 1992, ApJ, 395, L83
\re Okada, R., Fukue, J., \& Matsumoto, R. 1989, PASJ, 41, 133
\re Paczy\'nski, B. 1977, ApJ, 216, 822
\re Piran, T. 2000, Physics Reports, 333, 529
\re Sawada, K., Matsuda, T., \& Hachisu, I. 1986, MNRAS, 219, 75
\re Shakura, N. I., \& Sunyaev, R. A. 1973, A\&A, 24, 337
\re Shibata, M., \& Uryu, K. 2000, Phys. Rev., D61, 064001
\re Steeghs, D., Harlaftis, E. T., \& Horne, K. 1997, MNRAS, 290, L28
\re Stone, J. M., Pringle, J. E., \& Begelman, M. C. 1999, MNRAS, 310, 1002
\re Takahara, F. 1979, Prog. Theor. Phys., 62, 629
\re Thorne, K. S. 1987, in 300 years of Gravitation, ed. S. W. Hawking,
     \& W. Israel (Cambridge: Cambridge University Press)
\re Thorne, K. S. 1997, Rev. Mod. Astron., 10, 1
\re Tsubono, K., Fujimoto, M.-K. \& Kuroda, K. 1997,
    Gravitational Wave Detection (Tokyo: Universal Academy Press)
\re van Putten, M. H. P. M. 2001, Phys. Report, 345, 1
\re Woodward, J. W., Tohline, J. W., \& Hachisu, I. 1994, ApJ, 420, 247
\re Woosley, S. E. 1993, ApJ, 405, 273
\re Yamada, S., \& Sato, K. 1995, ApJ, 450, 245
\end{document}